# Temperature dependence of transport spin polarization in NdNi$_5$ measured using Point Contact Andreev reflection

Sourin Mukhopadhyay and Pratap Raychaudhuri[a]

Department of Condensed Matter Physics and Materials Science, Tata Institute of Fundamental Research, Homi Bhabha Rd, Colaba, Mumbai 400005, India.

and

Devang A Joshi and C V Tomy

Department of Physics, Indian Institute of Technology, Bombay, Powai, Mumbai 400076, Maharashtra, India

We report a study in which Point contact Andreev reflection (PCAR) spectroscopy using superconducting Nb tip has been carried out on NdNi$_5$, a ferromagnet with a Curie temperature of $T_C \sim 7.7K$. The measurements were performed over a temperature range of 2-9K which spans across the ferromagnetic transition temperature. From an analysis of the spectra, we show that (i) the temperature dependence of the extracted value of transport spin polarization closely follows the temperature dependence of the spontaneous magnetization; (ii) the superconducting quasiparticle lifetime shows a large decrease close to the Curie temperature of the ferromagnet. We attribute the latter to the presence of strong ferromagnetic spin fluctuations in the ferromagnet close to the ferromagnetic transition temperature.

72.00.00, 73.23.Ad, 72.80.Ga, 74.80.Fp, 75.50.-y

[a]electronic mail: pratap@tifr.res.in

# Introduction

With the emergence of "spintronics" [1] the ability to manipulate electronic spin as well the charge in a solid has gained particular significance. Central to the progress of this field is the identification of novel ferromagnetic materials with large degree of spin polarization which can be used as spin source in novel electronic devices, which will use both the charge and spin of the electrons. In this context experimental determination of the spin polarization (defined as $P = \frac{N_\uparrow(E_F) - N_\downarrow(E_F)}{N_\uparrow(E_F) + N_\downarrow(E_F)}$, where $N_\uparrow(E_F)$ and $N_\downarrow(E_F)$ are the spin up and spin down density of states at Fermi level) in ferromagnets has gained particular significance. Conventionally, two techniques have been widely used for the determination of spin polarization in ferromagnetic materials. The first one, pioneered by Meservey and Tedrow [2] involves spin dependent tunneling from a ferromagnet to a superconductor when the superconducting density of states is Zeeman split by the application of a magnetic field. This technique involves the tedious process of fabricating a tunnel junction and the need to apply a large perturbing magnetic field of several Telsa. The second one, spin resolved photoemission [3], relies on the measurement of spin of electrons that emerge from the surface of a ferromagnet close to the Fermi level. However, since the electrons measured in photoemission experiments emerge from a depth of few angstroms to few tens of angstroms from the surface this technique is extremely surface sensitive. More recently, a third technique, namely, Point Contact Andreev Reflection (PCAR) spectroscopy [4] has emerged as arguably one of the most popular probes [5,6,7,8,9,10,11,12] for the measurement of transport spin polarization in ferromagnetic materials. This technique relies on the fact that the process of Andreev reflection [13], through which an electron incident on a normal metal/superconductor (N/S) interface with energy less than the superconducting energy gap ($\Delta$) gets reflected back as a hole in the opposite spin band of the metal, is strongly suppressed when the normal metal electrode is a ferromagnet. The transport spin polarization is thus determined from the analysis the conductance (G(V)) versus voltage (V) characteristics of a ballistic or diffusive point contact established between a superconducting tip and the ferromagnet. This technique has several advantages: It is simple to implement, does not require fabrication of tunnel junctions and can be used for a wide variety of materials; it also does not require the application of a magnetic field.

One obvious limitation of the PCAR technique is that the temperature range of measurement is limited by the superconducting transition temperature of the superconducting tip. The determination of spin polarization of ferromagnets is thus typically carried out at temperatures which are two orders of magnitude lower than the ferromagnetic transition temperature ($T_C$) of most ferromagnets. Since much of the practical interest is in the value of spin polarization close to room temperature, the bulk spin polarization at elevated temperature is often estimated based on the assumption, $P(T) \propto M_s(T)$, namely, that the spin polarization of the electrons close the Fermi level is proportional to the spontaneous magnetization ($M_s(T)$) of the ferromagnet[14,15,16,17]. Though the spin polarization as a function of temperature has been measured using spin polarized photoemission for a few ferromagnets,[18,19] this simple relation has so far remained experimentally unverified even for relatively simple ferromagnets[20]. The experimental verification of this relation is important for two reasons. First, $M_s$ is a bulk property that depends on the total number difference of the up and down spin electrons, whereas P is only sensitive to the electrons close to Fermi level; this intuitive relation is thus based on a simplistic picture which is strictly valid only for free electron like parabolic bands. Second, both PCAR and Meservey-Tedrow technique measure the spin polarization in the transport current[21] ($P_t$) rather than the spin polarization in the density of states, which depends on a weighted average of the density of states and Fermi velocity of the two spin bands. The transport spin polarization is given by[21] $P_t = \dfrac{\langle N_\uparrow(k)(v_{k\uparrow})^n \rangle_{FS} - \langle N_\downarrow(k)(v_{k\downarrow})^n \rangle_{FS}}{\langle N_\uparrow(k)(v_{k\uparrow})^n \rangle_{FS} + \langle N_\downarrow(k)(v_{k\downarrow})^n \rangle_{FS}}$, where n=1 for a ballistic point contact, and n=2 for diffusive point contact. It has been shown the spin polarization measured from Meservey-Tedrow technique is identical to the diffusive point contact. It is not obvious that the temperature dependence of the spin polarization extracted from these techniques should follow $M_s$. However, since this relation forms the basis of much of the experimental studies related to spin transport in a variety of spintronic materials it is important to investigate if it holds good it in a typical ferromagnet.

In this paper, we report the PCAR studies carried out on $NdNi_5$, a ferromagnet[22,23,24] with a $T_C$~7.7K. Like most $RNi_5$ (R=rare-earth) compounds, the moment primarily reside on the Nd sites inducing a small moment on the Ni sites. The low $T_C$ of this material enables us to extract the transport spin polarization ($P_t$) all the way up to $T_C$ by carrying out the measurements using a finely cut superconducting Nb tip (superconducting transition temperature, $T_c$=9.2K). The central result of this paper is that the temperature dependence of $P_t$, in $NdNi_5$ closely follows the temperature dependence of $M_s$, validating the relation $P(T) \propto M_s(T)$. In addition, we observe a decrease of the superconducting quasiparticle lifetime close to the ferromagnetic transition temperature. We attribute this effect to the large spin fluctuation in the $NdNi_5$ close to the critical temperature. This hypothesis is further supported by PCAR measurements on 3 other systems: (i) The ferromagnetic metal, Fe, for which the $T_C$ much higher than the temperature range of measurements and therefore the effect of spin fluctuations are unimportant; and (ii) the non-ferromagnetic metals[25] $YbFe_4Sb_{12}$ and $CaFe_4Sb_{12}$ where detailed magnetic studies show the evidence of large ferromagnetic spin fluctuations at low temperatures.

**Experimental Details**

Polycrystalline sample $NdNi_5$ was prepared by repeated arc melting of the stoichiometric amounts of the constituent elements on water cooled copper hearth in a purified argon atmosphere. The button was flipped and re-melted several times to ensure the homogeneity. Titanium button was used as an oxygen getter. The total weight loss during the arc melting was less than 0.5% and hence the alloy compositions were assumed to remain unchanged from the original stoichiometric ratios. Room temperature powder X-ray diffraction pattern of the sample was obtained using Panalytical X-ray diffractometer equipped with Cu-Kα radiation. In order to obtain the lattice parameters of the compound and confirm its homogeneity to the accuracy of the X-ray pattern, a Rietveld refinement using the FullProf program of the obtained XRD pattern was done. The resistivity (ρ) and magnetization (M) of the sample was measured in the temperature range 3K-300K using a home made resistivity setup and Quantum Design SQUID magnetometer respectively. PCAR measurements were performed in the temperature range 2.4K to 9K in a continuous flow He4 cryostat. The sample was polished to a mirror finish and loaded immediately for experiment to avoid surface degradation. A mechanically cut sharp Nb tip was brought in contact with the sample at low temperatures using a differential screw arrangement and the conductance versus voltage characteristics of the contact was measured using a 4-probe current modulation technique. Typical contact resistance in these measurements ranged between 10-20$\Omega$.

**Results and Discussion**

Figure 1 shows the Rietveld refinement of the X-ray pattern of NdNi$_5$ which forms in a CaCu$_5$ type hexagonal structure with a space group P6/mmm. The excellent agreement with the experimental diffraction pattern (deduced from the near zero difference plot) confirms that the material is single phase. The obtained lattice parameters, $a$ = 4.953 Å and $c$ = 3.967 Å are in agreement with the published reports[22].

The magnetization versus temperature of the NdNi$_5$ sample (shown in Figure 2(a)) measured at 500 Oe reveals a sharp ferromagnetic transition with T$_C$~7.7K. The resistivity (*inset* Figure 2(a)) also shows a pronounced anomaly at the same temperature. Figure 2(b) shows the isothermal M-H curves recorded at various temperatures. The M-H curve does not saturate up to 2.5T due to the large magnetocrystalline anisotropy in this material. At low temperatures, M$_s$ was estimated by linearly extrapolating the high field slope of the M-H curve. The value of M$_s$ at 2K (~1.68$\mu_B$/f.u.), is much lower than the expected saturation moment of 3.28$\mu_B$ for the free Nd$^{3+}$ ion. This is due to crystal field splitting of the 4f energy levels in Nd as has been shown in numerous previous studies[26,27]. Above 6.6K the high field M-H curve was no longer linear. In this temperature range M$_s$ was estimated from the Arott plots (M$^2$ vs. H/M) shown in Figure 2(c).

In figure 3(a) we show the PCAR G(V)-V spectra (normalized with respect to the conductance values at large bias, $G_n$) for $NdNi_5$ recorded at various temperatures. The normalized conductance spectra are fitted with a modified Blonder-Tinkham-Klapwizk[28] (BTK) theory which takes into account the spin polarization of the ferromagnet[29]. Within this model the current through the point contact consists of a fully polarized ($I_p$) and an unpolarized ($I_u$) component such that the total current in terms of the transport spin polarization is given by, $I=P_t I_p+(1-P_t)I_u$. The unpolarized component of the current undergoes Andreev reflection in the same way as in an interface between a non magnetic metal and a superconductor. For the polarized component on the other hand the Andreev reflected hole cannot propagate and decays as an evanescent wave close to the N/S interface. $I_u$ and $I_p$ are thus calculated by using the BTK expression for the current,

$$I_{u,p}(V) \propto \int_{-\infty}^{\infty} [f(E-eV)-f(E)][1+A_{u,p}(E)-B_{u,p}(E)]dE, \quad (1)$$

where $f(E)$ is the Fermi function, and $A_u(E)$ ($B_u(E)$) and $A_p(E)$ ($B_p(E)$) are the Andreev reflection and normal reflection coefficients, calculated by solving the Bogolubov-de Gennes (BdG) equations for a non magnetic metal/superconductor and a fully polarized ferromagnet/superconductor respectively. To simulate a realistic interface, a delta function potential of the form $V_0\delta(x)$ is assumed at the interface. This delta function potential, parameterized within the model as a dimensionless parameter, $Z = \dfrac{V_0}{\hbar v_F}$, takes into account multiple effects: First, it takes into account the effect of any oxide barrier that may be present at the interface; second, Z also accounts for an effective barrier arising from the Fermi velocity mismatch between the normal metal and the superconductor. The lifetime of the superconducting ($\tau$) quasiparticle is incorporated in this model by including a broadening parameter[30] $\Gamma\ (=\frac{\hbar}{\tau})$ while solving the BdG equations. We have thus four fitting parameters: $P_t$, $\Delta$, $\Gamma$ and $Z$. In order to reduce the number of free parameters we restrict $\Delta$ to within 5% of its BCS value for Nb at all temperatures. The resulting fit of the spectra at various temperatures is shown through solid lines in Figure 3(a).

Figure 3(b) shows the extracted values of $P_t$ of NdNi$_5$ as a function of temperature. While, in the absence of a detailed estimate of the elastic and inelastic mean free paths, we cannot ascertain whether the contacts are in the ballistic or diffusive limit, the latter is more likely since our sample has a relatively small residual resistivity ratio; i.e. $\frac{\rho(300K)}{\rho(3K)} = 3.59$. In the same graph we also show the temperature variation of $M_s$. It can be easily seen that the temperature variation of the two quantities is similar. To further illustrate this point in the inset we plot $P_t$ as a function of $M_s$. Barring temperatures very close to $T_C$ where we see a small deviation the points fall on a straight line with zero intercept corroborating the relation, $P(T) \propto M_s(T)$.

We now focus our attention on the temperature dependence of the other quantities, namely, $\Delta$, $\Gamma$ and $Z$. The temperature variation of these three quantities extracted from the fits in Fig.3 (a) is shown in Figure 3(c). We would like to note that all the spectra could be fitted very well with the constraint on $\Delta$ stated earlier, with $\Delta(T=0)=1.45$meV. For temperatures in the range 2.4K to 4K the spectra can be fitted without incorporating any broadening parameter ($\Gamma$=0). Above 4K, $\Gamma$ gradually increases and reaches a maximum value of $\Gamma$=0.65meV at 8K which coincides with the $T_C$ of NdNi$_5$. Above 8K, $\Gamma$ decreases reaching a value of $\Gamma$=0.16meV at 9K. The barrier parameter, $Z$, on the other hand remains constant in the range 2.4K to 4K with a value of Z=0.27. Above 4K, Z increases monotonically up to 8K and tends to saturate to a value of Z=0.635.

Before discussing the implications of these results we would first like to comment on the reliability of the fits of the PCAR spectra, particularly at elevated temperatures. With increasing temperature, PCAR spectra get gradually thermally smeared. At temperatures greater than 8K the most dominant feature of the spectra, namely, the two peaks in the conductance spectra associated with the superconducting energy gap gets smeared into one broad peak. Since the saturation in the value of Z happens in this temperature range, the fit of the spectra for T>8K need careful attention. To cross-check the reliability of our fits in Figure 4 we show two fits of the same spectra taken at 8.5K: One fit with the parameters shown in Figure 3(b)-(c) (solid line) and the second one (dashed line) where Z is deliberately reduced and $\Gamma$ is adjusted to obtain the best possible fit. Though the parameters can be adjusted to reproduce the peak value in the normalized G(V) vs. V curves in both cases the latter does not reproduce the width of the curve close to zero bias (*inset* Figure 4). Nevertheless above 7.5K the uncertainty in the value of Z and $\Gamma$ significantly increases as shown in figure 3(c).

We now come to the significance of the temperature variation of $\Gamma$ and Z. First we focus on the temperature dependence of $\Gamma$. The increase in $\Gamma$, which peaks close to the critical point of the ferromagnet signifies a corresponding decrease in quasiparticle lifetime at the same temperature. It is known that ferromagnetic spin fluctuation in an s-wave superconductor increases the singlet state repulsion[31]. Since ferromagnetic spin fluctuation in NdNi$_5$ is maximum at temperatures close to T$_C$, it would be natural to attribute the decrease in the superconducting quasiparticle lifetime to the proximity of the superconductor to strong ferromagnetic spin fluctuation. We would however like to note that, a self consistent solution of this problem should however also take into account the corresponding decrease in Δ. At present we do not have a self consistent model to incorporate this effect into our analysis. The temperature variation of Z on the other hand is more complex to understand. It has been pointed out by several authors that in the analysis of a ferromagnet/superconductor interface, Z implicitly incorporates much more physics than a simple potential barrier at the interface. For a non-magnetic metal/superconductor interface, Z is given by,[32] $Z = Z_i + \frac{(r-1)^2}{4r}$, where $r$ is ratio of the Fermi velocity in the normal metal and the superconductor. The first term ($Z_i$) arises from a physical barrier arising from imperfect interface and oxide barrier and the second term incorporates the effect of Fermi velocity mismatch between the normal metal and the superconductor. In the case where the normal metal is a ferromagnet with different Fermi velocity of the up and down spin band, the derivation of the second term is not straightforward. In this case it is expected that the second term in Z would incorporate the effect of an average potential barrier experienced by the Fermi velocity mismatch between up and down spin electrons and the superconductor. As the temperature of the ferromagnet is raised towards the ferromagnetic transition, the Fermi velocities of the up ($v_{F\uparrow}$) and down ($v_{F\downarrow}$) spin bands gradually change due the reduction in exchange splitting and eventually become equal at $T_C$. The gradual change in Z from 4K to 8K and the leveling off to a constant value above 8K suggests that this evolution in $v_{F\uparrow}$ and $v_{F\downarrow}$ is reflected in the temperature dependence of Z. It is also expected that the ferromagnetic spin fluctuations will have additional effects on Z. This issue is currently beyond the scope of our paper ad needs to be explored theoretically.

In order to cross check the conjecture that spin fluctuations in the ferromagnet causes the decrease in the superconducting quasiparticle lifetime we have performed PCAR measurements on two different kinds of systems which are in two extremes in terms of spin fluctuations. The first measurement was on Fe (using a Nb tip), for which the $T_C \sim 1043K$ is two order of magnitude larger than the temperature range over which the measurement is carried out. Thus for Fe both ferromagnetic spin fluctuations as well as the decrease in exchange splitting is likely to be insignificant. The second systems are the filled skuterrudites[25], $CaFe_4Sb_{12}$ and $YbFe_4Sb_{12}$ which are nearly ferromagnetic metals for which large spin fluctuations are expected to be present even at the lowest temperatures. Figure 5 (a-c) shows the temperature dependence of the PCAR spectra and best fit parameters for the Fe-Nb point contact. As expected within experimental errors the transport spin polarization (Figure 5(b)) of Fe is constant over the entire temperature range of measurement with $P_t \sim 40\%$. Figure 5(c) shows the temperature variation of Z and $\Gamma$. Z is constant over the entire temperature range. $\Gamma$ on the other hand remains zero except at temperature very close to the transition temperature of the superconductor, where it shows a slight increase. This slight increase may reflect the intrinsic decrease of the quasiparticle lifetime of the superconductor as theoretically predicted,[33] and experimentally observed in strong coupling superconductors[34]. Figure 6(a) shows the PCAR spectra on the $CaFe_4Sb_{12}$. In this case the PCAR spectra recorded at 3.5K can be fitted only by incorporating a finite value of $\Gamma(=0.35meV)$. The situation is similar for $YbFe_4Sb_{12}$ (Figure 6(b)), where the PCAR spectra recorded at 2.3K can only be fitted incorporating $\Gamma=0.9meV$. Consistent with the large value of $\Gamma$, the superconducting energy gap is reduced from the bulk value for Nb, i.e. $\Delta=1.2meV$ for the Nb-

CaFe$_4$Sb$_{12}$ contact and Δ=0.85meV for the Nb-YbFe$_4$Sb$_{12}$ contact. Though for the extreme situation in YbFe$_4$Sb$_{12}$ where Γ>Δ, the standard BTK equations are not strictly applicable, this illustrates the effect of spin fluctuations on the quasiparticle lifetime in a PCAR experiment. Our results are consistent with earlier reports on CaFe$_4$Sb$_{12}$ and YbFe$_4$Sb$_{12}$ where it has been shown from magnetization and heat capacity that spin fluctuations are larger in the latter compound.

## Conclusions

In summary, we have shown that in NdNi$_5$, the temperature dependence of the transport spin polarization extracted from PCAR spectroscopy closely follows the temperature variation of the spontaneous magnetization M$_s$. We have also shown that the superconducting quasiparticle lifetime extracted by fitting the PCAR spectra show a minimum close to the ferromagnetic transition temperature of NdNi$_5$. Through a detailed comparison with measurements carried out on the ferromagnet Fe and the nearly ferromagnetic compounds CaFe$_4$Sb$_{12}$ and YbFe$_4$Sb$_{12}$, we attribute this decrease in the quasiparticle lifetime to the effect of large spin fluctuations close to the critical temperature of the ferromagnet. We believe that a detailed theoretical understanding of the role of spin fluctuations could establish PCAR as an alternative technique to probe to investigate ferromagnetic spin fluctuations in nearly ferromagnetic metals.

## Acknowledgements

We thank Y. Grin and Andreas Leithe-Jasper for providing high quality polycrystalline samples of CaFe$_4$Sb$_{12}$ and YbFe$_4$Sb$_{12}$, Igor Mazin and S. K. Dhar for helpful discussions and John Jesudasan, Vivas Bagwe and Subhash Pai for technical help. SM would like to thank the TIFR Endowment Fund for partial financial support.

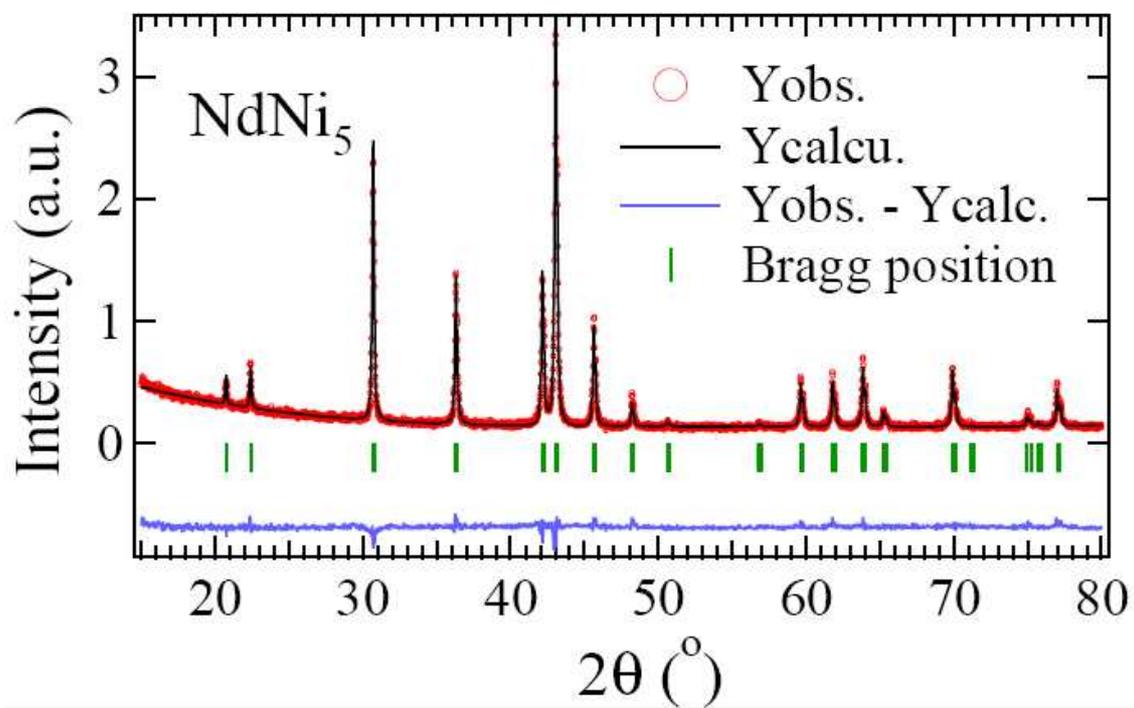

**Figure 1:** The observed and calculated X-ray diffraction pattern of NdNi$_5$ along with their difference. The vertical bars indicate Bragg reflections.

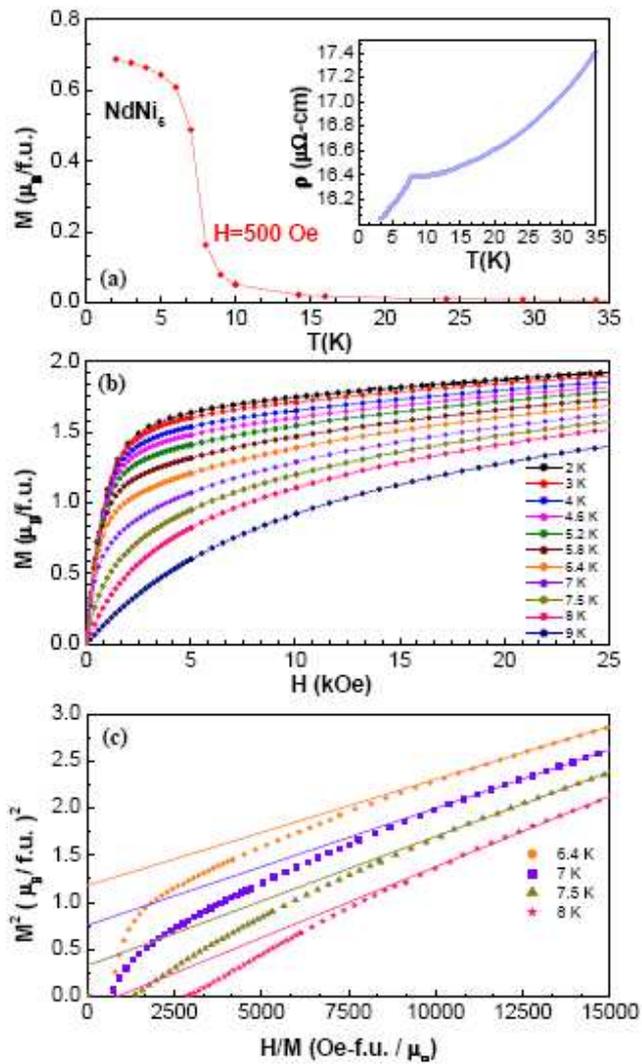

**Figure 2:** (a) Magnetisation versus temperature of NdNi$_5$ measured in 500Oe. The *inset* shows the resistivity as a function of temperature. (b) Isothermal M-H curves at various temperatures. (c) Arott plots (M$^2$ versus H/M) at four temperatures close to T$_C$.

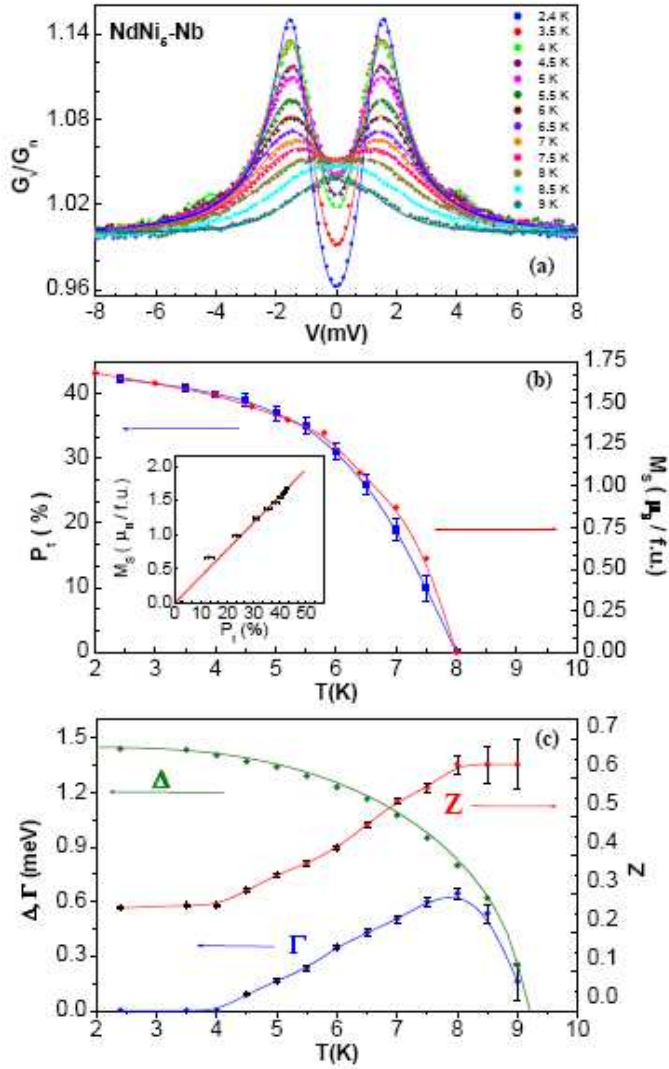

**Figure 3:** (a) Normalized PCAR spectra of NdNi$_5$-Nb point contact at different temperatures. The solid lines show the fits to the spectra using the modified BTK model. (b) Temperature variation of the $P_t$ (squares) and $M_s$ (circles). The solid lines are guide to the eye. The *inset* shows the linear variation of $M_s$ with $P_t$. (c) Temperature dependence of the superconducting energy gap ($\Delta$), the barrier parameter (Z) and the broadening parameter ($\Gamma$). The solid line passing through $\Delta$ shows the expected temperature variation from BCS theory. The solid lines passing through Z and $\Gamma$ are guide to the eye.

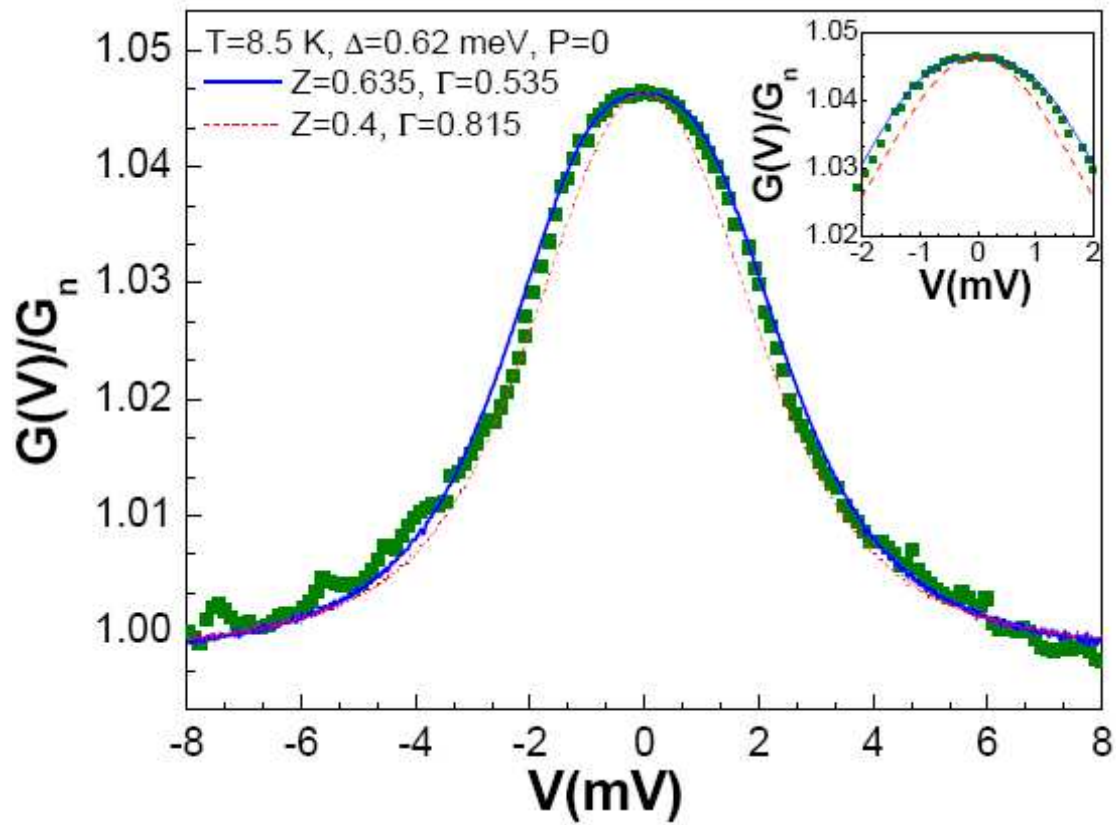

**Figure 4: Comparison of the fits of PCAR spectra for Nb-NdNi$_5$ contact at 8.5K with two sets of fitting parameters shown in solid and dashed lines. The fit parameters for the two fits are shown in the figure. The *inset* is an enlarged view for the low bias region.**

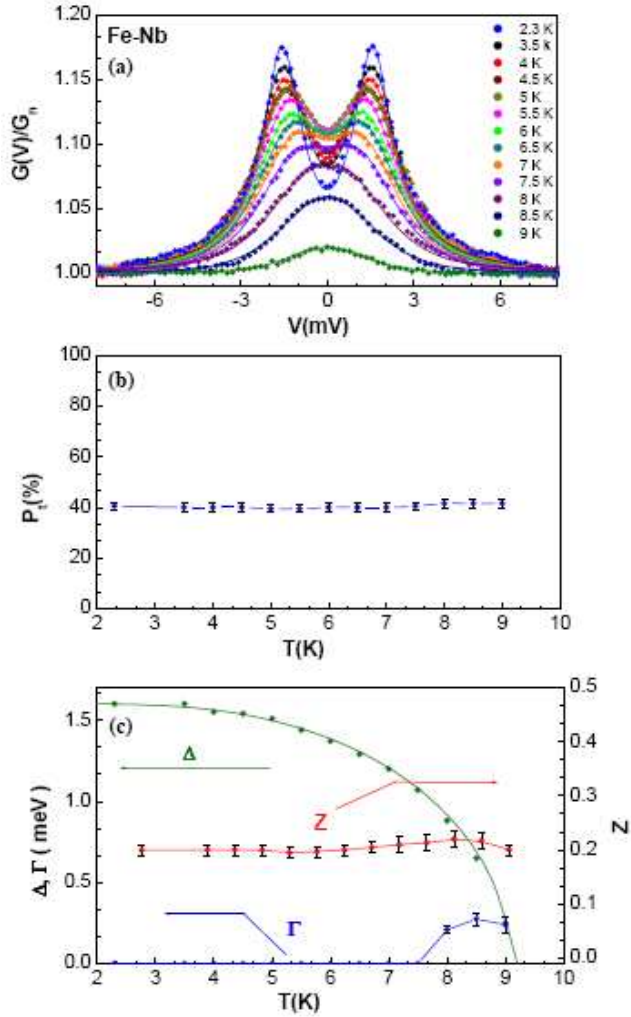

**Figure 5:** (a) Normalized PCAR spectra of Fe-Nb point contact at different temperatures. The solid lines show the fits to the spectra using the modified BTK model. (b) Temperature variation of the transport spin polarization ($P_t$) in the temperature range 2.3K to 9K. The solid line is a guide to the eye. (c) Temperature dependence of the superconducting energy gap ($\Delta$), the barrier parameter (Z) and the broadening parameter ($\Gamma$). The solid line passing through $\Delta$ shows the expected temperature variation from BCS theory. The solid lines passing through Z and $\Gamma$ are guide to the eye.

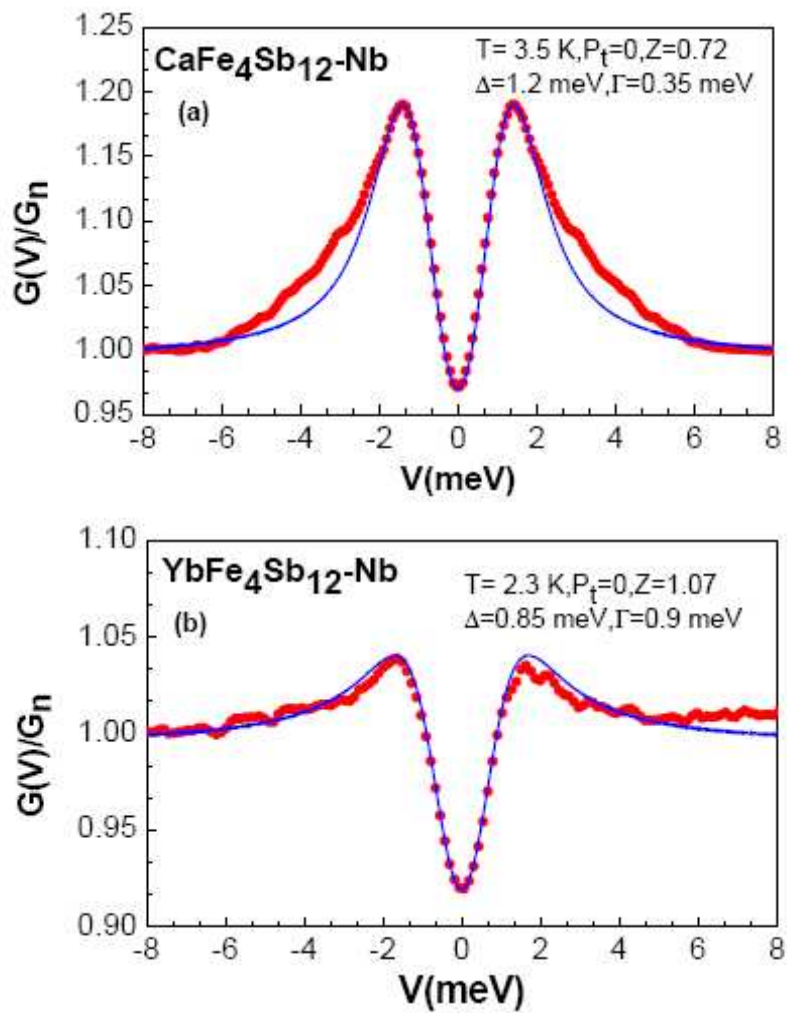

**Figure 6:** Normalized PCAR spectra for (a) CaFe$_4$Sb$_{12}$-Nb point contact at 3.5K; and (b) YbFe$_4$Sb$_{12}$-Nb point contact at 2.3K. The solid lines are the best fits of the spectra to the modified BTK model. The best fit parameters are also shown in the figures.